\documentclass[journal,twoside,web]{ieeecolor}
\usepackage{generic}
\usepackage{cite}
\usepackage{amsmath,amssymb,amsfonts}
\usepackage{algorithmic}
\usepackage{graphicx}
\usepackage{textcomp}
\definecolor{ieeeblue}{RGB}{0,67,147}
\def\BibTeX{{\rm B\kern-.05em{\sc i\kern-.025em b}\kern-.08em
    T\kern-.1667em\lower.7ex\hbox{E}\kern-.125emX}}
\begin{document}
\title{Characterization and Modeling of Silicon-on-Insulator Lateral Bipolar Junction Transistors at Liquid Helium Temperature}
\author{Yuanke Zhang, Yuefeng Chen, Yifang Zhang, Jun Xu, Chao Luo, and Guoping Guo
	\thanks{This work was supported by the National Natural Science Foundation of China (No. 12034018), Innovation Program for Quantum Science and Technology (No. 2021ZD0302300). \emph{ (Yuanke Zhang and Yuefeng Chen contributed equally to this work.)} (Corresponding author: Chao Luo, e-mail: lc0121@ustc.edu.cn)}    
	\thanks{The authors are with University of Science and Technology of China (USTC), Hefei 230026, Anhui, China, and also with CAS Key Lab of Quantum Information, Hefei 230026, Anhui, China.} 
}   

\maketitle

\begin{abstract}
Conventional silicon bipolars are not suitable for low-temperature operation due to the deterioration of current gain ($\beta$). In this paper, we characterize lateral bipolar junction transistors (LBJTs)  fabricated on silicon-on-insulator (SOI) wafers down to liquid helium temperature (4 K). The positive SOI substrate bias could greatly increase the collector current and have a negligible effect on the base current, which significantly alleviates $\beta$ degradation at low temperatures. We present a physical-based compact LBJT model for 4 K simulation, in which the collector current ($\textit{I}_\textbf{C}$) consists of the tunneling current and the additional current component near the buried oxide (BOX)/silicon interface caused by the substrate modulation effect. This model is able to fit the Gummel characteristics of LBJTs very well and has promising applications in amplifier circuits simulation for silicon-based qubits signals.
\end{abstract}

\begin{IEEEkeywords}
Cryogenic, lateral bipolar junction transistors, silicon-on-insulator, characterization, modeling, tunneling, substract modulation
\end{IEEEkeywords}

\section{Introduction}
\label{sec:introduction}
\IEEEPARstart {C}{ryogenic} electronics has a promising application for deep aerospace exploration, neutrino physics experiments, infrared focal plane array surfaces, etc., and has been studied to design and implement the manipulation and readout circuits of quantum bits (qubits) in recent years\cite{b1,b2,b3,b4,b5,b6,b7,b8,b9,b35}. Bipolar junction transistors (BJTs) with high current gain ($\beta$) have been widely used as low-noise local signal amplifiers and can be a potential candidate for spin readout devices of semiconductor qubits\cite{b10,b11,b12,b13,b14}. Therefore, heterojunction bipolar transistors are widely studied due to their useful amplification performance in a wide temperature range even down to millikelvin\cite{b10,b11,b12}. However, large-scale quantum computing requires the integration of a large number of qubits and circuits on a single chip. In order to be compatible with the fabrication process of silicon-based qubits, silicon homojunction BJTs remain the most promising candidate. Unfortunately, due to the carrier freeze-out in the base region and the narrowing of the bandgap associated with the emitter, $\beta$ degrades severely with decreasing temperature in conventional silicon bipolars\cite{b15,b16,b17,b18}.

\par To overcome this problem, fabricating the homojunction BJT laterally on a silicon-on-insulator substrate provides a promising solution\cite{b14,b19,b20,b21}. With a voltage applied to the SOI substrate, an additional current component can be generated near the buried oxide (BOX)/silicon interface. In n-p-n type symmetric LBJTs, a positive SOI substrate voltage ($V_{\rm BOX}$) could significantly increase the collector current ($I_{\rm C}$) with almost no change in the base current ($I_{\rm B}$), and thus increase the current gain. Moreover, previous studies have shown that the modulation effect of $V_{\rm BOX}$ remains effective at low temperatures and the signal-to-noise ratio (SNR) gain can also be ameliorated by adjusting $V_{\rm BOX}$\cite{b14}, which demonstrates the potential application of LBJTs in amplifying weak electronic signals generated at cryogenic temperatures. In order to design cryogenic circuits based on LBJTs, an accurate compact simulation model is necessary. However, compact modeling of LBJTs at low temperatures remains unexplored.

\par In this article, the low-temperature characteristics of LBJTs fabricated on SOI wafers ranging from 300 K to 4 K are presented. For the first time, a physical-based LBJT compact model is proposed for 4 K simulation. The collector current is consist of the tunneling current and the additional drift-diffusion current component caused by the positive SOI substrate bias. The model calculation results show very good agreement with the measurement data of LBJTs, especially the modulation effect of $V_{\rm BOX}$.

\par This article is organized as follows. In Section II, we provide a description of the device structure and the cryogenic measurement setup. Section III describes characterization of the devices from 300 K to 4 K and discusses the cryogenic behaviors. In Section IV, we present a physics-based LBJT model for 4 K simulation, and finally, we conclude this article in Section V.

\section{Experimental Details}
The schematic of an n-p-n type LBJT fabricated on SOI wafers is shown in Fig. 1(a). Two different sizes of LBJTs are tested in this paper: emitter length ($L_{\rm E}$)-base width ($W_{\rm B}$)-emitter wing widths ($W_{\rm E}$)-collector wing widths ($W_{\rm C}$) = 10-0.1-0.2-0.2 and 5-0.1-0.15-0.2 $\mu$m. More detailed fabrication information can be referred to elsewhere\cite{b20}. The measurement setup is shown in Fig. 1(b)-(c). The diced sample chips are bonded to the chip carriers using aluminum (Al) wires [Fig. 1(b)] and the electrical characteristic measurement is performed by a Keysight B1500A semiconductor analyzer. The low-temperature environment is provided by liquid nitrogen (77 K)/helium (4 K) dewar [Fig. 1(c)]. A dip-stick with a rhodium-iron resistance thermometer is placed at different heights inside the dewar to reach temperatures between 300 K and 77 K/4 K and it is pre-placed for 15 minutes at each temperature to ensure measurement environment stability. Differential $\beta$ = d$I_{\rm C}$/d$I_{\rm B}$ is used in this paper.

\begin{figure}[!h]
	\centerline{\includegraphics[width=\columnwidth]{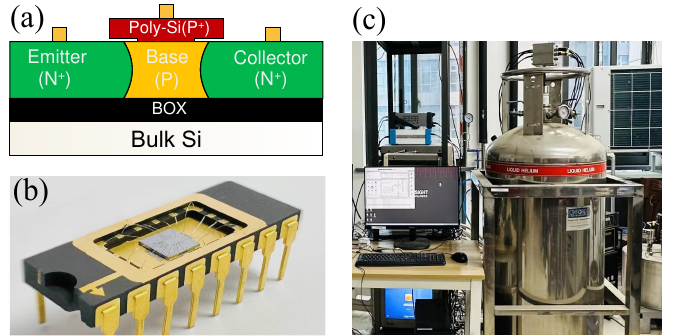}}
	\caption{(a) Schematic cross-sectional view of a symmetric n-p-n LBJT. (b) Sample chip, wire-bonded to a chip carrier with Al-wire bonds. (c) Liquid helium dewar with a dip-stick inside.}
	\label{fig1}
\end{figure}

\section{Characterization}

The Gummel characteristics of the LBJTs measured under $V_{\rm BOX}$ = 0 V and $V_{\rm BOX}$ = 12 V at various temperatures are shown in Fig. 2(a) and (b), respectively. Throughout the article, both $I_{\rm C}$ and $I_{\rm B}$ are normalized by emitter length ($L_{\rm E}$). The slope of $I_{\rm C}$ increases with decreasing temperature due to a $kT/q$ dependence. It should be noted that $I_{\rm C}$-$V_{\rm BE}$ curves measured at 20 K and 4 K essentially overlap under $V_{\rm BOX}$ = 0 V [Fig. 2(a)] and the overlap disappears with a positive $V_{\rm BOX}$ = 12 V [Fig. 2(b)]. This phenomenon can be attributed to two different current transport mechanisms of LBJTs: the E-C tunneling current inside the LBJT and the drift-diffusion current near the BOX/silicon interface. At low temperatures, the potential barrier in the base region prevents the injection of electrons, and $I_{\rm C}$ is mainly composed of the E-C tunneling current\cite{b10,b14,b22}. Due to the saturation of electron temperature, $I_{\rm C}$ is almost independent of temperature when $T$$\leq$17 K\cite{b22}, and thus the overlapping phenomenon occurs. Differently, an additional depletion region near the BOX/silicon interface is generated under a positive $V_{\rm BOX}$ and thus the drift-diffusion current between the collector and emitter is enhanced. As the drift-diffusion current is temperature dependent, the overlap disappears under $V_{\rm BOX}$ = 12 V, as shown in Fig. 2(b).

\begin{figure}[!h]
	\centerline{\includegraphics[width=3.1in]{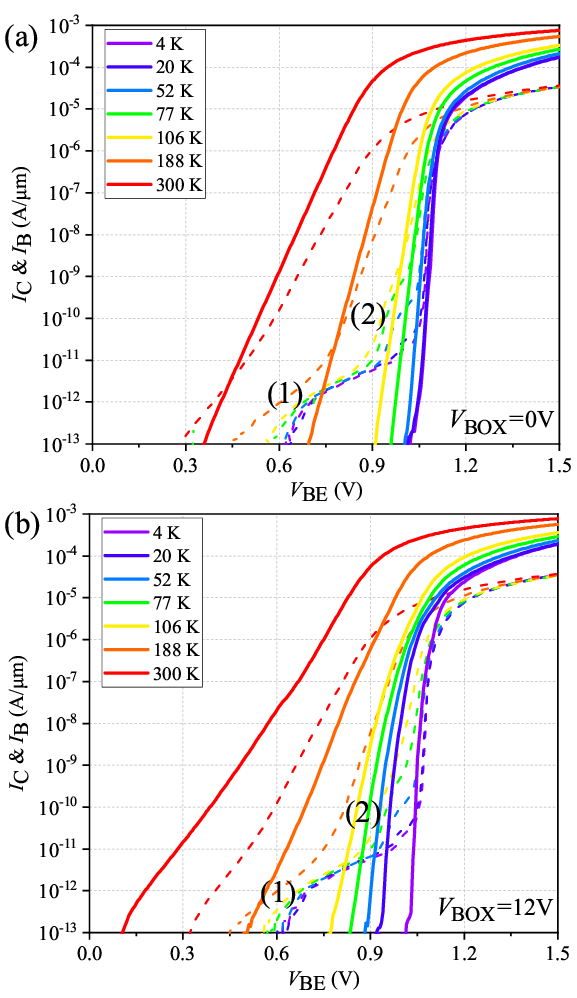}}
	\caption{$I_{\rm C}$ (solid lines) and $I_{\rm B}$ (dash lines) versus $V_{\rm BE}$ of LBJT with $L_{\rm E}$-$W_{\rm B}$-$W_{\rm E}$-$W_{\rm C}$ = 10-0.1-0.2-0.2 $\mu$m at different temperatures under (a) $V_{\rm BOX}$ = 0 V; (b) $V_{\rm BOX}$ = 12 V, $V_{\rm CE}$ = 1 V.}
	\label{fig2}
\end{figure}

\par $I_{\rm C}$ and $I_{\rm B}$ versus $V_{\rm BE}$ of LBJTs with two different sizes under $V_{\rm BOX}$ = 0$\sim$12 V are shown in Fig. 3. Due to the additional drift-diffusion current regulated by the $V_{\rm BOX}$, $I_{\rm C}$ is significantly enhanced with increasing $V_{\rm BOX}$ under medium $V_{\rm BE}$ values. With further increase of $V_{\rm BE}$ ($V_{\rm BE}$$\textgreater$0.8 V at 300 K and $V_{\rm BE}$$\textgreater$1.1 V at 4 K), the current transport inside the LBJTs (i.e. the traditional BJT transport at 300 K and tunneling at 4 K) plays a dominant role and the influence of $V_{\rm BOX}$ is not noticeable anymore. Moreover, due to the shorter poly-Si lines and lower base resistance, the LBJT with smaller $L_{\rm E}$ delivers a higher $I_{\rm C}$\cite{b20}. Surprisingly, $I_{\rm B}$ is negligibly affected by $V_{\rm BOX}$. The transport of $I_{\rm B}$ is mainly concentrated on the upper surface of LBJTs, hence $V_{\rm BOX}$ can hardly affect the injection barrier of holes from the base to the emitter. The increased $I_{\rm C}$ and the unaffected $I_{\rm B}$ under the modulation effect of $V_{\rm BOX}$ imply an improvement in $\beta$, as shown in Fig. 4(a) and (b). Under $V_{\rm BOX}$ = 12 V, $\beta$ (at $I_{\rm B}$ = 1 nA/$\mu$m) is improved by $\sim$10 times and $\sim$10$^{3}$ times at 300 K and 4 K, respectively. 

\par In addition, $\beta$ versus $I_{\rm B}$ at different temperatures under $V_{\rm BOX}$ = 0 V and 12 V are shown in Fig. 4(c)-(d). As expected, $\beta$ deteriorates with decreasing temperature and is significantly improved by the positive SOI substrate bias. The curves at 4  K and 20 K are very similar in Fig. 4(c), which can be attributed to the overlap of $I_{\rm C}$ discussed above. Due to the reduction of the diffusion coefficient ($D_{\rm B}$), $\beta$ under high injection conditions ($I_{\rm B}$$\textgreater$10$^{-6}$ A/$\mu$m) reduces with decreasing temperature at each temperature, as shown in Fig. 4(c) and (d).
 
\begin{figure}[!h]
	\centerline{\includegraphics[width=\columnwidth]{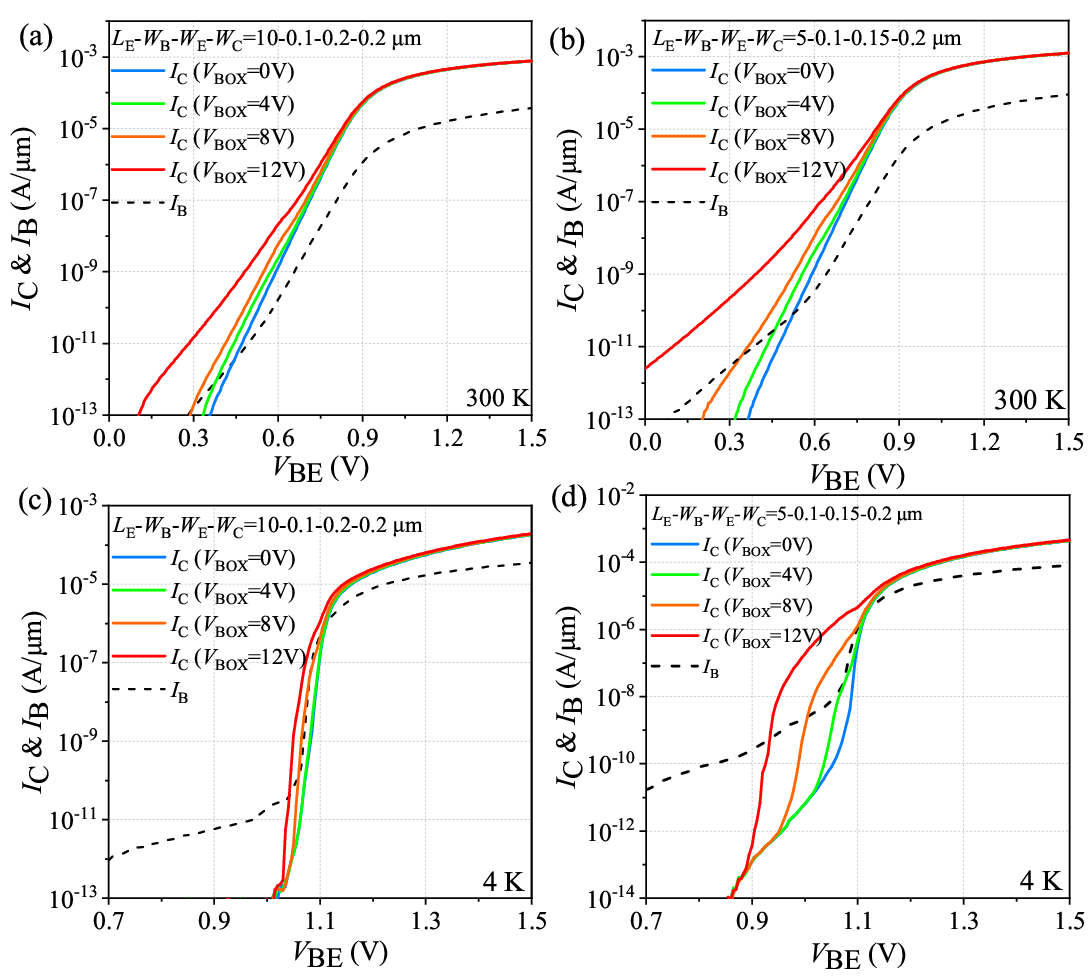}}
	\caption{$I_{\rm C}$ (solid lines) and $I_{\rm B}$ (dash lines) versus $V_{\rm BE}$ of two different sizes of LBJTs, (a)-(b) at 300 K; (c)-(d) at 4 K, $V_{\rm CE}$ = 1 V. $V_{\rm BOX}$ changes from 0 V to 12 V in steps of 4 V.}
	\label{fig3}
\end{figure}

\begin{figure}[!h]
	\centerline{\includegraphics[width=\columnwidth]{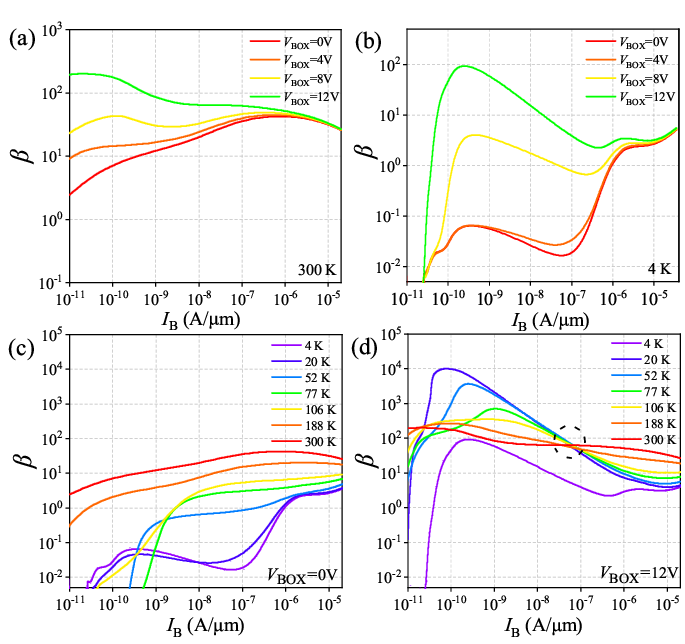}}
	\caption{$\beta$ versus $I_{\rm B}$ of the LBJTs with $L_{\rm E}$-$W_{\rm B}$-$W_{\rm E}$-$W_{\rm C}$ = 10-0.1-0.2-0.2 $\mu$m at (a) 300 K; (b) 4 K under different $V_{\rm BOX}$, and under (c) $V_{\rm BOX}$ = 0 V; (d) $V_{\rm BOX}$ = 12 V at different temperatures. }
	\label{fig4}
\end{figure}

\section{Compact Modeling}
As $I_{\rm B}$ is composed of diode currents from the base-emitter (B-E) and base-collector (B-C) junctions, its mechanism remains consistent from room temperature to low temperatures. The Gummel-Poon (GP) bipolar model \cite{b23} is used to describe $I_{\rm B}$ characteristics of LBJT in this work. With $V_{\rm BE}$ ranging from 0 to 1.5 V and $V_{\rm CE}$ remaining 1 V, the B-C junction is always in reverse bias or weak bias. Therefore, the current of the B-C junction is negligible and $I_{\rm B}$ can be written in the form\cite{b34}

\begin{equation}
I_{\rm B}=A_{\rm E} \frac{q n_i^2}{G_{\rm E}} \exp \left(\frac{q V_{\rm BE}}{kT}\right)
\end{equation}

\noindent where $A_{\rm E}$ is the area of the emitter-base junction, $n_{i}$ is the intrinsic carrier density, and $G_{\rm E}$ is the emitter Gummel number, which is inversely proportional to the diffusion coefficient $D_{\rm B}$. However, $D_{\rm B}$ and $n_{i}$ reduce dramatically with decreasing temperature. At 4 K, $n_{i} \approx$ 10$^{-678}$ cm$^{-3}$, which lies outside the range of IEEE double-precision arithmetic (10$^{-308}$$\sim$10$^{308}$)\cite{b24}, thus resulting in the parameter $I_{\rm SE}$ (B-E leakage saturation current) in the GP model being too small for computers to calculate. Therefore, we modify Eq. (1) as a summation of the diffusion current and the recombination current

\begin{equation}\begin{split}
I_{\rm{B}}&=\frac{I_{\rm{S}}}{B_{\rm{f}}} \{[\exp (S_{\rm diff}(V_{\rm BE}-V_{\rm diff}))-1]\\&+  I_{\rm{SE}}[\exp (S_{\rm RE}(V_{\rm BE}-V_{\rm RE}))-1]\}\cdot f_{\rm fermi}
\end{split}\end{equation}

\noindent where $S_{\rm diff}$ and $S_{\rm RE}$ are slope parameters of $I_{\rm B}$-$V_{\rm BE}$ in semi-logarithmic scale. $V_{\rm diff}$ and $V_{\rm RE}$ are the voltages corresponding to diffusion and recombination conductance exceeding the minimum conductance across each nonlinear device (GMIN) in SPICE\cite{b25}. $I_{\rm S}$, $I_{\rm SE}$, and $B_{\rm f}$ represent the modified saturation current coefficient, the B-E leakage saturation current coefficient, and the ideal forward maximum gain, respectively. $f_{\rm fermi}$ is used to guarantee a zero current at a zero $V_{\rm BE}$\cite{b27,b28}. Moreover, the base parasitic resistance $R_{\rm B}$ is also taken into account to precisely calculate $I_{\rm B}$ in the large injection region and the Newton-Raphson iteration\cite{b26} is used for solving the current and voltage of the intrinsic base. 
\par As we discussed in Sec. III, $I_{\rm C}$ is mainly composed of the E-C tunneling current at low temperatures. Assuming that the potential barrier in the base region is parabolic in shape, the tunneling current $I_{\rm T\_tunl}$ is given by\cite{b29}

\begin{equation}
I_{\rm T\_tunl}=A_{1}\sqrt{v_b}\left[\frac{\exp \left(a_{1} v_e / \sqrt{v_b}\right)-1}{a_{1} v_e / \sqrt{v_b}}-1\right] \exp \left(-a_{1} \sqrt{v_b}\right)
\end{equation}

\noindent when $v_b\geq v_e$. And for $v_b<v_e$ condition, $I_{\rm T\_tunl}$ is given by

\begin{equation}\begin{split}
I_{\rm T\_tunl}&=A_{1}\sqrt{v_b}\left[\left[\exp\left(a_{1}\sqrt{v_b}\right)-1\right]\left(1-\frac{v_b}{v_e}\right)\right.\\&\left.+\frac{\exp\left(a_{1}\sqrt{v_b}\right)-1}{a_{1}v_e/\sqrt{v_b}}-\frac{v_b}{v_e}\right] \exp \left(-a_{1} \sqrt{v_b}\right)
\end{split}\end{equation}

\noindent where $v_b = 1- V_{\rm BE}/V_{\rm DEi}$ and $v_e = \Delta W_{\rm E}/qV_{\rm DEi}$. $A_{1}$, $a_{1}$, and $V_{\rm DEi}$ present the current density prefactor, exponent factor, and the built-in voltage of the internal BE junction, respectively.  $\Delta W_{\rm E}$ is the parameter related to the height of the potential barrier. It is worth noting that the Fermi distribution function $f(\rm E)$ at $T$ = 0 K is used in the solution of $I_{\rm T\_tunl}$. At 0 K, $f(\rm E)$ is a step function, thus such concise Eq. (3) and (4) are obtained. Although the difference in $f(\rm E)$ may lead to some deviations, it is acceptable at the target application temperature of our model, i.e., 20-100 mK (integration with qubits) or 1-4 K (integration with qubits controller).
\par As $V_{\rm BE}$ increases further, the emitter carrier energy will approach or even exceed the potential barrier height. The tunneling current $I_{\rm T\_tunl}$ will tend to level off or even decrease, and the hot carrier transmission current ($I_{\rm T\_hc}$) plays a dominant role, which is given by

\begin{equation}
I_{\rm T\_hc}=A_{1}a_{1}\frac{v_e}{2}\left(1-\frac{v_b}{v_e}\right)^2
\end{equation}

\begin{figure}[!h]
	\centerline{\includegraphics[width=\columnwidth]{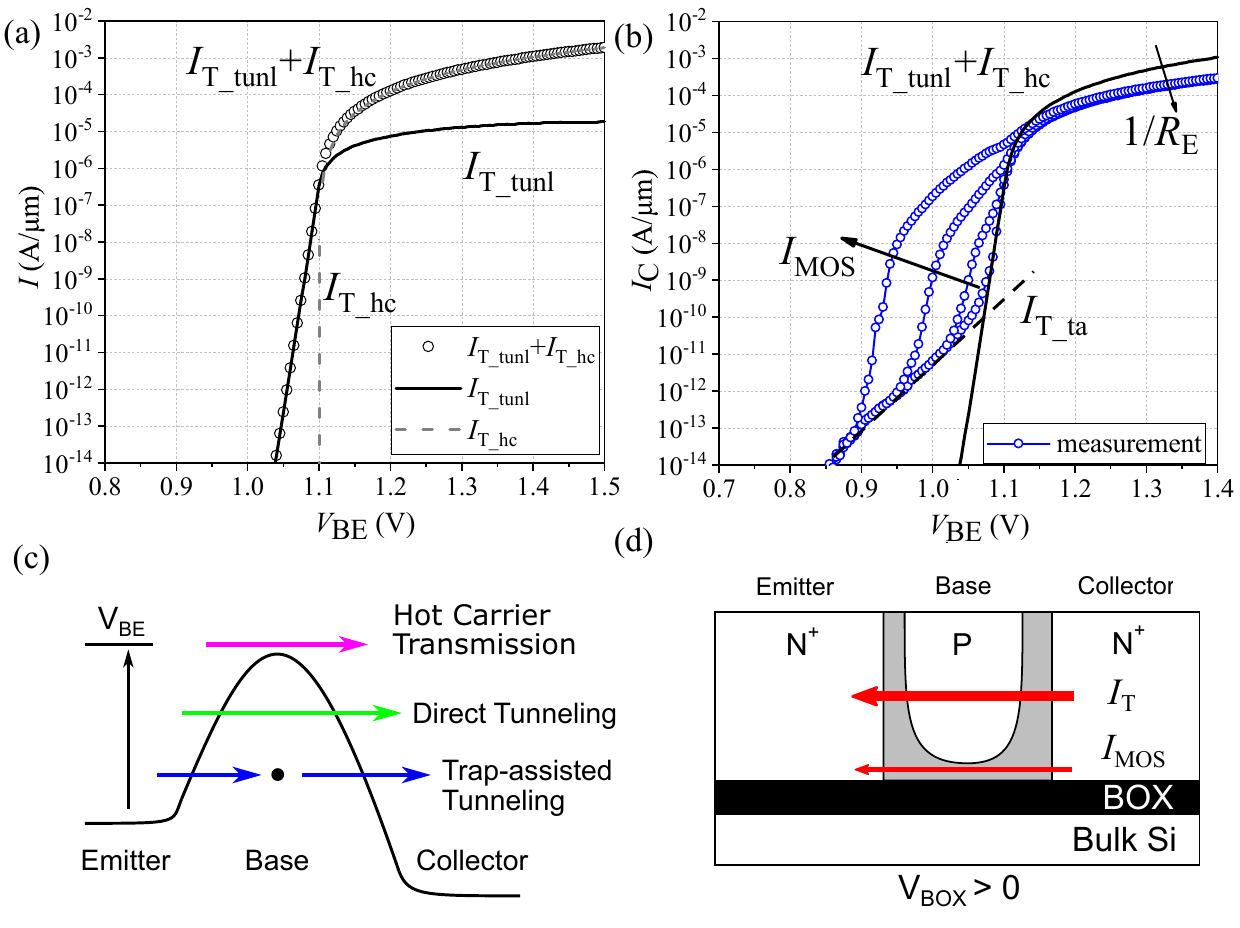}}
	\caption{(a) $I_{\rm T\_tunl}$ and $I_{\rm T\_hc}$ versus $V_{\rm BE}$. (b) The current components of the LBJT compact model. The measruement result is the same as the data in Fig. 3(d). (c) Qualitative illustration of the tunneling mechanisms under different $V_{\rm BE}$ values. (d) Schematic cross section showing the current transport in LBJTs with positive $V_{\rm BOX}$ values.}
	\label{fig5}
\end{figure}

\noindent where the parameter definitions are the same as Eq. (3) and (4). Fig. 5(a) shows the contribution of $I_{\rm T\_tunl}$ and $I_{\rm T\_hc}$ in our model, in which $A_{1}$ = 8$\times$10$^{-5}$ A/$\mu$m, $a_{1}$ = 300, $V_{\rm DEi}$ = 1.7 V, and $\Delta W_{\rm E}$ = 0.601 eV. In addition, in order to accurately describe the $I_{\rm C}$ behavior at low $V_{\rm BE}$ values, the trap-assisted tunneling current $I_{\rm T\_ta}$\cite{b31,b32} is also taken into account in this model and simply described by an exponential function [see $I_{\rm T\_ta}$ in Fig.5 (b)]. When there is a limited distribution of traps in the bandgap of the base, carriers can tunnel from the emitter to the collector with the assistance of the traps. Therefore, the total tunneling current can be given by $I_{\rm T}$=$I_{\rm T\_tunl}$+$I_{\rm T\_hc}$+$I_{\rm T\_ta}$ and a qualitative illustration of the tunneling mechanisms under different $V_{\rm BE}$ values is shown in Fig. 5(c).

\begin{figure}[!h]
	\centerline{\includegraphics[width=3.1in]{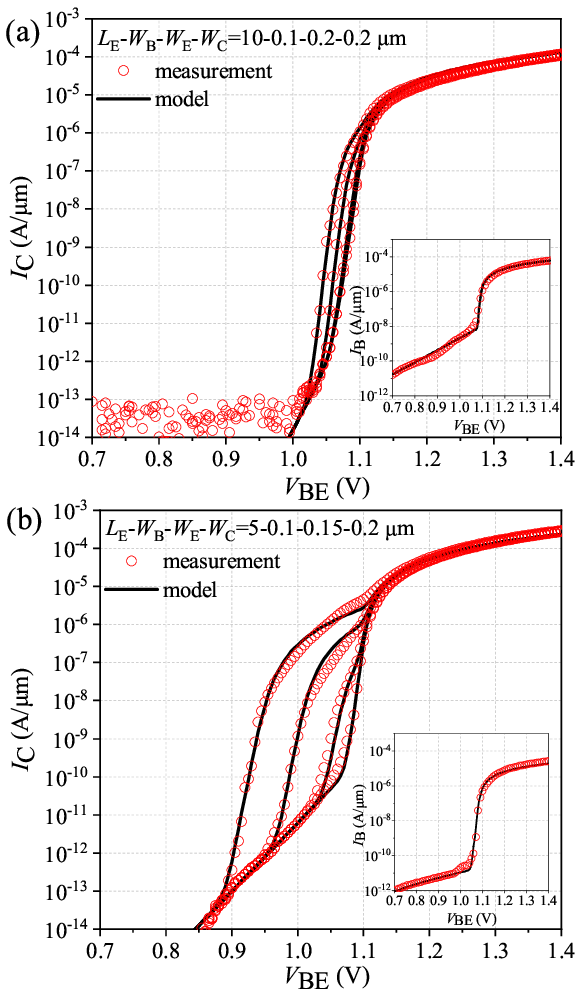}}
	\caption{$I_{\rm C}$ and $I_{\rm B}$ (inset) versus $V_{\rm BE}$ of the LBJTs measured (symbol) and calculated (solid line) at liquid helium temperature with $L_{\rm E}$-$W_{\rm B}$-$W_{\rm E}$-$W_{\rm C}$ = 10-0.1-0.2-0.2 (a) and 5-0.1-0.15-0.2 $\mu$m (b) on $V_{\rm BOX}$ from 0 up to 12 V by 4 V steps.}
	\label{fig5}
\end{figure}

\par When a positive $V_{\rm BOX}$ is applied, the base region near the BOX is partially depleted and an additional drift-diffusion current is generated near the BOX/silicon interface, as qualitatively illustrated in Fig. 5(d). In this case, a symmetrical LBJT can be viewed as an upside-down MOSFET. The emitter and collector correspond to the source (S) and drain (D), BOX corresponds to the gate (G) oxide, and the base corresponds to the silicon substrate (sub). Differently, the gate voltage ($V_{\rm G}$) in MOSFETs is applied to a poly-Si gate rather than to the bulk Si in LBJTs. To modify this deviation, we introduce an effective gate voltage $V_{\rm G\_eff} = a_2 V_{\rm BOX}$ in the model, and applying $V_{\rm BE}$ is equivalent to the modulation of $V_{\rm sub}$ in MOSFETs. To calculate this MOSEFT-like drift-diffusion current $I_{\rm MOS}$, the overdrive voltage $V_{\rm ov}$ = $V_{\rm G}-V_{\rm TH}$ in commercial MOSFET model \cite{b30} is replaced by

\begin{equation}
V_{\rm ov} = V_{\rm G\_eff}-V_{\rm TH}-\gamma\left(\sqrt{2\phi_B-V_{BE}}-\sqrt{2\phi_B}\right)
\end{equation}

\noindent where $V_{\rm TH}$ the threshold voltage, $\phi_B$ is the bulk Fermi potential, and $\gamma$ is the body effect parameter in MOSFETs. The effect of the emitter parasitic resistance $R_{\rm E}$ is also calculated by the Newton-Raphson iterative method\cite{b26}. Therefore, the total $I_{\rm C}$ can be given by 

\begin{equation}
I_{\rm C} = I_{\rm T} + I_{\rm MOS}
\end{equation}

\noindent and take the $I_{\rm C}$-$V_{\rm BE}$ characteristics of the LBJT with $L_{\rm E}$-$W_{\rm B}$-$W_{\rm E}$-$W_{\rm C}$ = 5-0.1-0.15-0.2 $\mu$m for example, the contribution of each current component in $I_{\rm C}$ is shown in Fig. 5(b). The parameter-fitting results of the proposed compact model for $L_{\rm E}$-$W_{\rm B}$-$W_{\rm E}$-$W_{\rm C}$ = 10-0.1-0.2-0.2 and  5-0.1-0.15-0.2 $\mu$m LBJTs at 4 K are shown in Fig. 6(a) and (b), respectively. Good matching of the measurement and calculation results is obtained in both devices and the proposed model is ready to use for LBJT-contained cryogenic circuit design.

\section{Conclusion}
In this article, we present the characterization and modeling of LBJTs fabricated on SOI wafers at liquid helium temperature. At low temperatures, $I_{\rm C}$ is mainly composed of the E-C tunneling current and the MOSEFT-like drift-diffusion current generated by positive SOI substrate bias. Based on the modeling of the two current components above, a physical-based LBJT compact model is proposed for 4 K simulation and it shows good fitting results with the measurement data. The proposed model can be used to design and simulate the LBJT-contained cryogenic circuits for local quantum signal amplification.

\section*{Acknowledgement}
The device fabrication was done by Prof. Zhen Zhang's group in the Ångström Microstructure Laboratory (MSL) at Uppsala University. Dr. Qitao Hu, Dr. Si Chen, Prof. Zhen Zhang are acknowledged for the device design and fabrication, and the technical staff of MSL are acknowledged for their process support.


\begin{thebibliography}{00}

\bibitem{b1} A. Ruffino, T.-Y Yang, J. Michniewicz, Y. Peng, E. Charbon, and M. F. Gonzalez-Zalba, ``A cryo-CMOS chip that integrates silicon quantum dots and multiplexed dispersive readout electronics", \emph{Nat. Electron.}, vol. 5, Jan. 2022, pp. 53-59, doi: \textcolor{ieeeblue}{10.1038/s41928-021-00687-6.}

\bibitem{b2} E. Charbon,  F. Sebastiano,  A. Vladimirescu,  H. Homulle,  S. Visser, L. Song, and R. M. Incande, ``Cryo-CMOS for quantum computing,'' in \emph{IEDM  Tech. Dig.}, Dec. 2016, pp. 13–15, doi: \textcolor{ieeeblue}{10.1109/IEDM.2016.7838410.}

\bibitem{b3} M. H. Devoret and R. J. Schoelkopf, ``Amplifying quantum signals with the single-electron transistor,” \emph{Nature}, vol. 406, no. 6799, pp. 1039–1046, Aug. 2000, doi:\textcolor{ieeeblue}{10.1038/35023253.}

\bibitem{b4} F. A. Zwanenburg, A. S. Dzurak, A. Morello, M. Y. Simmons, L. C. L. Hollenberg, G. Klimeck, S. Rogge, S. N. Coppersmith, and M. A. Eriksson, ``Silicon quantum electronics,” \emph{Rev. Modern Phys.}, vol. 85, no. 3, pp. 961–1019, Jul. 2013, doi: \textcolor{ieeeblue}{10.1103/RevModPhys.85.961.}

\bibitem{b35} E. Gutiérrez-D, J. Deen, and C. Claeys, \emph{Low Temperature Electronics: Physics, Devices, Circuits, and Applications}. San Diego, CA, USA: Academic, Oct. 2001, doi: \textcolor{ieeeblue}{10.1016/B978-0-12-310675-9.X5000-2.}

\bibitem{b5} B. Patra, R. M. Incandela, J. P. G. van Dijk, H. A. R. Homulle, L. Song, M. Shahmohammadi, R. B. Staszewski, A. Vladimirescu, M. Babaie, F. Sebastiano, and E. Charbon, ``Cryo-CMOS circuits and systems for quantum computing applications,” \emph{IEEE J. Solid-State Circuits}, vol. 53, no. 1, pp. 309–321, Jan. 2018, doi: \textcolor{ieeeblue}{2018. 10.1109/JSSC.2017.2737549.}

\bibitem{b6} J. R. Hoff, G. W. Deptuch, Guoying Wu, and Ping Gui, ``Cryogenic Lifetime Studies of 130 nm and 65 nm nMOS Transistors for High-Energy Physics Experiments," \emph{ IEEE Trans. Nucl. Sci.}, vol. 62, no. 3, pp. 1255–1261, 2015, \textcolor{ieeeblue}{doi: 10.1109/TNS.2015.2433793.}

\bibitem{b7} B. Patra, ``CMOS circuits and systems for cryogenic control of silicon quantum processors," 2021. [Online]. Available: https://doi.org/10.4233/uuid:cea59727-fda2-41e1-ba87-9404ef22202d.

\bibitem{b8} Y. Zhang et al., ``Characterization and modeling of native MOSFETs down to 4.2 K,” \emph{IEEE Trans. Electron Devices}, vol. 68, no. 9, pp. 4267–4273, Sep. 2021, doi: \textcolor{ieeeblue}{10.1109/TED.2021.3099775.}

\bibitem{b9} Y. Liu, L. Lang, Y. Chang, Y. Shan, X. Chen and Y. Dong, ``Cryogenic Characteristics of Multinanoscales Field-Effect Transistors,” \emph{IEEE Trans. Electron Devices}, vol. 68, no. 2, pp. 456-463, 2021, doi: \textcolor{ieeeblue}{10.1109/TED.2020.3041438.}

\bibitem{b10} H. Ying, B. R. Wier, J. Dark, N. E. Lourenco, L. Ge, A. P. Omprakash, M. Mourigal, D. Davidovic, and J. D. Cressler, ``Operation of SiGe HBTs down to 70 mK,” \emph{IEEE Electron Device Lett.}, vol. 38, no. 1, pp. 12–15, Jan. 2017, doi: \textcolor{ieeeblue}{10.1109/LED.2016.2633465.}

\bibitem{b11} D. Davidovic, H. Ying, J. Dark, B. R. Wier, L. Ge, N. E. Lourenco, A. P. Omprakash, M. Mourigal, and J. D. Cressler, ``Tunneling, current gain, and transconductance in silicon-germanium heterojunction bipolar transistors operating at millikelvin temperatures,”\emph{Phys. Rev. Appl.}, vol. 8, no. 2, Aug. 2017, Art. no. 024015,
doi: \textcolor{ieeeblue}{10.1103/physrevapplied.8.024015.}

\bibitem{b12} M. J. Curry, T. D. England, N. C. Bishop, G. Ten-Eyck, J. R. Wendt, T. Pluym, M. P. Lilly, S. M. Carr, and M. S. Carroll, ``Cryogenic preamplification of a single-electron-transistor using a silicon-germanium heterojunction-bipolar-transistor,” \emph{Appl. Phys. Lett.}, vol. 106, no. 20, May 2015, Art. no. 203505, doi: \textcolor{ieeeblue}{10.1063/1.4921308.}

\bibitem{b13} I. T. Vink, T. Nooitgedagt, R. N. Schouten, L. M. K. Vandersypen, and W. Wegscheider, ``Cryogenic amplifier for fast real-time detection of single-electron tunneling,” \emph{Appl. Phys. Lett.}, vol. 91, no. 12, Sep. 2007, Art. no. 123512, doi:\textcolor{ieeeblue}{ 10.1063/1.2783265.}

\bibitem{b14} S. Chen, C. Luo, Y. Zhang, J. Xu, Q. Hu, Z. Zhang, and G. Guo, ``Current gain enhancement for silicon-on-insulator lateral bipolar junction transistors operating at liquid-helium temperature,” \emph{IEEE Electron Device Lett.}, vol. 41, no. 6, pp. 800–803, June 2020, doi: \textcolor{ieeeblue}{10.1109/LED.2020.2985674.}

\bibitem{b15} J. D. Cressler, J. H. Comfort, E. F. Crabbe, G. L. Patton, J. M. C. Stork, J. Y.-C. Sun, and B. S. Meyerson, ``On the profile design and optimization of epitaxial Si- and SiGe-base bipolar technology for 77 k applications. I. transistor DC design considerations,” \emph{IEEE Trans. Electron Devices}, vol. 40, no. 3, pp. 525–541, Mar. 1993, doi: \textcolor{ieeeblue}{10.1109/16.199358.}

\bibitem{b16} H. Homulle, L. Song, E. Charbon, and F. Sebastiano, ``The cryogenic temperature behavior of bipolar, MOS, and DTMOS transistors in standard CMOS,”\emph{IEEE J. Electron Devices Soc.}, vol. 6, pp. 263–270, 2018, doi: \textcolor{ieeeblue}{10.1109/JEDS.2018.2798281.}

\bibitem{b17} L. Song, H. Homulle, E. Charbon and F. Sebastiano, ``Characterization of bipolar transistors for cryogenic temperature sensors in standard CMOS,” in \emph{2016 IEEE SENSORS}, 2016, pp. 1-3, doi:\textcolor{ieeeblue}{ 10.1109/ICSENS.2016.7808759.}

\bibitem{b18} H. Homulle, ``Cryogenic Electronics for the Read-Out of Quantum  Processors,” PhD thesis, Technical Univ. Delf, 2019, doi:\textcolor{ieeeblue}{10.4233/uuid:e833f394-c8b1-46e2-86b8-da0c71559538.}

\bibitem{b19} Q. Hu, S. Chen, S.-L. Zhang, P. Solomon, and Z. Zhang, ``Effects of substrate bias on low-frequency noise in lateral bipolar transistors fabricated on Silicon-on-Insulator substrate,” \emph{IEEE Electron Device Lett.}, vol. 41, no. 1, pp. 4–7, Jan. 2020, doi: \textcolor{ieeeblue}{10.1109/LED.2019.2953362.}

\bibitem{b20} Q. Hu, X. Chen, H. Norstrom, S. Zeng, Y. Liu, F. Gustavsson, S.-L. Zhang, S. Chen, and Z. Zhang, ``Current gain and low-frequency noise of symmetric lateral bipolar junction transistors on SOI,” in\emph{Proc. 48th Eur. Solid-State Device Res. Conf. (ESSDERC)}, Sep. 2018, pp. 258–261, doi:\textcolor{ieeeblue}{ 10.1109/ESSDERC.2018.8486918.}

\bibitem{b21} J.-B Yau, J. Cai, and T. H. Ning, ``Substrate-Voltage Modulation of Currents in
Symmetric SOI Lateral Bipolar Transistors,” \emph{IEEE Trans. Electron Devices}, vol. 63, no. 5, pp. 1835–1839, May 2016, doi: \textcolor{ieeeblue}{10.1109/TED.2016.2543528.}

\bibitem{b33} R. S. Müller and T. I. Kamins, \emph{Device Electronics for Integrated Circuits}, 3rd ed. New York, NY, USA: Wiley, 2003.

\bibitem{b22} H. Rucker, J. Korn, and J. Schmidt, ``Operation of sige HBTs at cryogenic temperatures,” in\emph{Proc. IEEE Bipolar/BiCMOS Circuits Technol. Meeting (BCTM)}, Oct. 2017, pp. 17–20, doi: \textcolor{ieeeblue}{10.1109/BCTM.2017.8112902.}

\bibitem{b23} F. Sischka,  ``GUMMEL-POON BIPOLAR MODEL,” Agilent Technologies, Munich, 1990). [Online]. Available: http://www.idea2ic.com/PlayWithSpice/pdf/G{\%}20U{\%}20M{\%}20M{\%}20E{\% }20L{\%}20-{\%}20P{\%}20O{\%}20O{\%}20N.pdf

\bibitem{b34} T. Yuan and T. H. Ning, \emph{Fundamentals of modern VLSI devices}, Cambridge university press, 2021.

\bibitem{b24} A. Beckers, F. Jazaeri, and C. Enz, ``Cryogenic MOS transistor model,” \emph{IEEE Trans. Electron Devices}, vol. 65, no. 9, pp. 3617-3625, Sept. 2018, doi: \textcolor{ieeeblue}{10.1109/TED.2018.2854701.}

\bibitem{b25} ``Spectre Circuit Simulator Reference, Product Version 19.1" Cadence Design Systems, Inc., Jan. 2020.

\bibitem{b26} W. J. McCalla,  ``Fundamentals of computer-aided circuit simulation,” Springer Science $\&$ Business Media, 1987, doi:\textcolor{ieeeblue}{10.1007/978-1-4613-2011-1.}

\bibitem{b27} Y.-K Lin, J. P. Duarte, P. Kushwaha, H. Agarwal, H.-L Chang, A. Sachid, S. Salahuddin, and C. Hu, ``Compact Modeling Source-to-Drain Tunneling in Sub-10-nm GAA FinFET With Industry Standard Model,” \emph{IEEE Trans. Electron Devices}, vol. 64, no. 9, pp. 3576-3581, Sept. 2017, doi: \textcolor{ieeeblue}{10.1109/TED.2017.2731162.}

\bibitem{b28} L. Zhang and M. Chan, ``SPICE modeling of double-gate tunnel-FETs including channel transports,” \emph{IEEE Trans. Electron Devices}, vol. 61, no. 2, pp. 300–307, Feb. 2014, doi:\textcolor{ieeeblue}{10.1109/TED.2013.2295237.}

\bibitem{b29} M. Schröter and X. Jin, ``A Physics-Based Analytical Formulation for the Tunneling Current Through the Base of Bipolar Transistors Operating at Cryogenic Temperatures,” \emph{IEEE Trans. Electron Devices}, vol. 70, no. 1, pp. 247-253, Jan. 2023, doi:\textcolor{ieeeblue}{10.1109/TED.2022.3223885.}

\bibitem{b31} A. J. Joseph, J. D. Cressler and D. M. Richey, ``Operation of SiGe heterojunction bipolar transistors in the liquid-helium temperature regime,” \emph{IEEE Electron Device Lett.}, vol. 16, no. 6, pp. 268-270, June 1995, doi: \textcolor{ieeeblue}{10.1109/55.790731.}

\bibitem{b32} Y. Hanbin, ``Collector current transport mechanisms in SiGe HBTs operating at cryogenic temperatures,” PhD thesis, Georgia Institute of Technology, 2019, [Online]. Available: https://smartech.gatech.edu/handle/1853/61292.

\bibitem{b30} M. Bucher, C. Lallement, C. Enz, F. Théodoloz, and F. Krummenacher, ``The EPFL-EKV MOSFET model equations for simulation model version 2.6,” Dept. Electron. Lab, Swiss Federal Inst. Technol., Lausanne, Switzerland, Jun. 1997. [Online]. Available: https://www.epfl.ch/labs/iclab/wp-content/uploads/2019/02/ekv$\_$v262.pdf.
\end{thebibliography}
\end{document}